\documentstyle[aps,amssymb]{revtex}

\begin{document}
\title{Quantum non-demolition measurement of nonlocal variables and its application
in quantum authentication}
\author{Guo-Ping Guo, Chuan-Feng Li\thanks{%
Electronic address: cfli@ustc.edu.cn } and Guang-Can Guo\thanks{%
Electronic address: gcguo@ustc.edu.cn }}
\address{Laboratory of Quantum Communication and Quantum Computation and Department\\
of Physics, University of Science and Technology of China, Hefei 230026,\\
People's Republic of China}
\maketitle

\begin{abstract}
Quantun non-demolition (QND) variables are generlized to the nonlocal ones
by proposing QND measurement networks of Bell states and multi-partite GHZ
states, which means that we can generate and measure them without any
destruction. One of its prospective applications in the quantum
authentication system of the Quantum Security Automatic Teller Machine
(QSATM) which is much more reliable than the classical ones is also
presented.

PACS number(s): 03.65.Ud, 03.67.Dd
\end{abstract}

\section{Introduction}

In the 1970's, Braginsky, Thorme, Unruh, Caves and others introduced firstly
the concept of quantum non-demolition (QND) measurement in which a
measurement strategy is chosen that evades the undesirable effect of back
action\cite{r0}. This was to respond the problem caused by Heisenberg's
uncertainty relations. Using the fact that the quantum formalism describes
physical quantities as non-commuting operators ( that is, as mathematical
objects $A,$ $B$ such that $AB\neq BA$), the Heisenberg inequalities state
that the product of the dispersions of (the `uncertainty' in) $A$ and $B$
has a lower bound: $\vartriangle A\vartriangle B\geqslant \frac 12\left|
\left\langle AB-BA\right\rangle \right| $. Therefore, for non-commuting
operators, a very precise measurement of $A$, resulting in a very small
dispersion $\vartriangle A,$will be associated with a large value of $%
\vartriangle B$. This measurement back action has far-reaching consequences
from a practical point of view, since it may prevent the retrieval of the
initial result in a series of repeated measurements\cite{r1}. With the
development of the technical quality of optical sources and detectors, and
techniques of nonlinear optics, the emerging field of quantum optics seemed
to be particularly well suited for implementing QND measurements.

The basic requirement of QND measurement is the availability of the QND
variable which may be measured repeatedly giving predictable results for a
system. If a variable $A(t)$ (in the interaction picture) satisfying $[A(t),$
$A(t^{^{\prime }})]=0$, then the system in an eigenstate of $A(t_0)$ will
remain in this eigenstate for all subsequent times although the eigenvalues
may change. Such observable is called QND variable. Furthermore, this QND
variable must commute with the interaction Hamiltonian coupling the detector
and the meter or $\left[ \hat{H}_{inter},\text{ }A(t)\right] $ $=0$. The
soul of the quantum non-demolition measurement is to measure QND variable in
their eigenstates: if the system is already in eigenstate then repeated
measurements will cause no demolition to it and give predictable result; if
the system is not in the eigenstate, then after the measurement it is set to
eigenstate and keep in this eigenstate in the following repeated
measurements.

But until now, all QND observables considered are local. And it is well
known that nonlocality and its most celebrated manifestation form ---
entanglement is so important and fancy that it becomes one of the footstones
of the quantum mechanics. Myriads of attention has been attracted since it
was first noted by Einstein-Podolsky-Rose (EPR) and Schr\"{o}dinger. And the
experimental prove of this nonlocality confirms the validity of quantum
mechanics. Its famous embodiment, the EPR states $\Phi ^{\pm }=\frac 1{\sqrt{%
2}}(|11\rangle \pm |00\rangle ),$ $\Psi ^{\pm }=\frac 1{\sqrt{2}}(|10\rangle
\pm |01\rangle $, proposed by Bohm, was shown by Bell, which have stronger
correlation than that allowed by any local hidden variable theory.
Furthermore this kind of correlation of the quantum mechanics cannot be
generated by local operations and classical communications (LQCC). Here
local operations include unitary transformations, additions of ancilla,
measurements and throwing away parts of the system, all performed locally by
one party on his subsystem. Classical communication between parties is
included because it allows for the creation of mixed states that are
classically correlated but exhibit no quantum correlations. For
multi-partite system, there is also entanglement, such as generalized GHZ
states $|\Psi ^{\pm }(x_0,x_1,...,x_{n-1})\rangle $\cite{r2}, where $%
x_i=\{0,1\}$, which are much more complicated and less well understood even
today. Recently it has been realized that quantum resources can be useful in
information processing where quantum entanglement plays a key role in many
such application as quantum teleportation\cite{r3}, superdense coding\cite
{r4}, entanglement enhanced classical communication\cite{r5}, quantum error
correction\cite{r6}, quantum key distribution\cite{r7}, quantum
computational speedups\cite{r8}, quantum distributed computation\cite{r9},
and entanglement enhanced communication complexity\cite{r10}. Among these
applications, the common difficulty is to measurement these orthogonal
entangled bases such as $n$-particle GHZ states. So it may be very expedient
if we can measure these entanglement states without demolition it.

In this article we generalize the QND variables to the nonlocal ones by
proposing a novel QND measurement network of Bell states and multi-partite
GHZ states. In section II, the logic network of Bell operator QND
measurement is presented. In section III, the general logic network for the
generalized GHZ states is shown. And in the last section, one of its
potential applications in quantum authentication (QA) system of the Quantum
Security Automatic Teller Machine (QSATM) is proposed.

\section{QND Measurement of Bell Bases}

We firstly introduce some basic definitions and notations.

The universal two-bit Controlled-not gate $\Lambda (\hat{U})$ maps Boolean
values $\left( x,\text{ }y\right) $to $\left( x,\text{ }x\oplus y\right) ,$
where $\hat{U}=$ $\left( 
\begin{array}{cc}
0 & 1 \\ 
1 & 0
\end{array}
\right) .$ Let $\hat{H}$ denotes Hadamard gate $\frac 1{\sqrt{2}}\left( 
\begin{array}{cc}
1 & 1 \\ 
1 & -1
\end{array}
\right) $, which transforms state $|1\rangle (|0\rangle )$ into $\frac 1{%
\sqrt{2}}(|1\rangle +|0\rangle )(\frac 1{\sqrt{2}}(|1\rangle -|0\rangle ))$.
It can be realized by an optical half-wave plate(HWP) with its fast axis
oriented at $\frac \pi 8$ relative to the horizontal direction in quantum
optics.

The $n$-partite generalized GHZ bases in the Hilbert space $H_2^{\otimes n}$%
could be written as $2^{n-1}$ pairs of orthogonal states $\{|\Psi ^{\pm
}(x_0,x_1,...,x_{n-1})\}$. Define the phase qubit represents the sign
information between the pairs. And define part-parity bit represents the
parity of the border upon parties. For example, if the two parties are $%
|11\rangle $ or $|00\rangle $, they have even parity, otherwise they have
odd parity. It is obvious there are independent $(n-1)$ parity qubits for $n$%
-partite GHZ bases such as part-parity of partite 1 and 2, of partite 2 and
3, and so on. In addition, we define global-parity qubit if both terms in a
state contain even or odd number of $|1\rangle $. Not all states has
global-parity qubit information.

Let $a=\pm 1,$ and $a^{\prime }=\pm 1$ denote 2 possible outcomes of 2
possible measurements on the first qubits and similarly $b=\pm 1$ and $%
b^{\prime }=\pm 1$ for the second qubit. And associates the first
measurement the Pauli matrix $\hat{a}\hat{\sigma}$ with normalized 3-dim
vectors $\hat{a},$ and similarly for the other measurements. Then Bell
operator $B_2=$ $\hat{a}\hat{\sigma}\otimes \hat{b}\hat{\sigma}+$ $\hat{a}%
\hat{\sigma}\otimes \hat{b}^{\prime }\hat{\sigma}+$ $\hat{a}^{\prime }\hat{%
\sigma}\otimes \hat{b}\hat{\sigma}-$ $\hat{a}^{\prime }\hat{\sigma}\otimes 
\hat{b}^{\prime }\hat{\sigma}$, and similarly the $n$-particle Bell
operators $B_n=$ $B_{n-1}\otimes \frac 12\left( \hat{a}_n\hat{\sigma}+\hat{a}%
_n^{\prime }\hat{\sigma}\right) +B_{n-1}^{\prime }\otimes \frac 12\left( 
\hat{a}_n\hat{\sigma}-\hat{a}_n^{\prime }\hat{\sigma}\right) ,$ whose
eigenstates are Bell states and generalized GHZ states.\cite{N.H.} Obviously
these nonlocal operators $B_n\left| \Psi \right\rangle =\Psi \left| \Psi
\right\rangle $ are constant and can be taken as the QND variables. If we
can measure Bell states and generalized GHZ states without any destruction,
then the interaction Hamiltonian$\hat{H}_{inter}$ satisfies $\hat{H}%
_{inter}\left| \Psi \right\rangle =\left| \Psi \right\rangle .$ Then $\left[ 
\hat{H}_{inter},\text{ }B_n\right] $ $=0$ , and any states can be measured
predictably in the subsequent measurements.

Now consider the QND measurement of the Bell states, the logic network is
shown in Fig. 1, where the ancillary qubits: ancilla 3 and ancilla 4 are
initially set in state $|0\rangle $. The measurement process for the four
Bell bases: $\Phi ^{\pm }=\frac 1{\sqrt{2}}(|11\rangle \pm |00\rangle ),$ $%
\Psi ^{\pm }=\frac 1{\sqrt{2}}(|10\rangle \pm |01\rangle )$ can be divided
into four steps.

\begin{quote}
(1) The two particles 1 and 2 in Bell state to be measured come into two
paths with controlled-not gate $\Lambda (\hat U_1)$ and $\Lambda (\hat U_2)$
separately. Then particle 1 and 2 act as control bit in turn to manipulate
the ancilla 3:
\end{quote}

\begin{quotation}
\begin{eqnarray}
&&\Lambda (\hat U_1)\Lambda (\hat U_2)(\Phi _{12}^{\pm }\otimes |0\rangle _3)
\label{3} \\
&=&\Lambda (\hat U_2)\frac 1{\sqrt{2}}(|11\rangle _{12}\otimes |1\rangle
_3\pm |00\rangle _{12}\otimes |0\rangle _3)  \nonumber \\
&=&\frac 1{\sqrt{2}}(|11\rangle _{12}\pm |00\rangle _{12})\otimes |0\rangle
_3  \nonumber
\end{eqnarray}
\begin{eqnarray}
&&\Lambda (\hat U_1)\Lambda (\hat U_2)(\Psi _{12}^{\pm }\otimes |0\rangle _3)
\label{4} \\
&=&\Lambda (\hat U_2)\frac 1{\sqrt{2}}(|10\rangle _{12}\otimes |1\rangle
_3\pm |01\rangle _{12}\otimes |0\rangle _3)  \nonumber \\
&=&\frac 1{\sqrt{2}}(|10\rangle _{12}\pm |01\rangle _{12})\otimes |1\rangle
_3  \nonumber
\end{eqnarray}
\end{quotation}

Seen from the equations above, states $\Phi ^{\pm }$ and $\Psi ^{\pm }$ set
ancilla 3 to state $|0\rangle _3$ and $|1\rangle _3$ respectively and keep
themselves unchanged. We see that the information of the parity bit is
extracted out by the ancilla 3 bit.

(2) Particle 1 and 2 are transformed by a pair of Hadamard gates $\hat{H}_1$
and $\hat{H}_2$ locally: 
\begin{equation}
\hat{H}_1\hat{H}_2\left( 
\begin{array}{c}
\Phi ^{+} \\ 
\Phi ^{-} \\ 
\Psi ^{+} \\ 
\Psi ^{-}
\end{array}
\right) =\left( 
\begin{array}{c}
\Phi ^{+} \\ 
\Psi ^{+} \\ 
\Phi ^{-} \\ 
-\Psi ^{-}
\end{array}
\right) .  \label{5}
\end{equation}

We can see the transformation is just a unitary transformation among the
four bases. But it is very important for the information of the phase qubit
is translated into the parity qubit by this transformation.

(3) As in step 1, the ancilla 4 pick out the parity qubit information
translated from the phase qubit in step 2 by the controlled not gates. Then
the four Bell bases are discriminated (see Table 1).

\[
\text{Table 1: The states of the ancillas corresponding to each Bell state} 
\]
\[
\begin{tabular}{|lllll|}
\hline
\multicolumn{1}{|l|}{$\Psi _{12}$} & \multicolumn{1}{l|}{$\Psi ^{+}$} & 
\multicolumn{1}{l|}{$\Psi ^{-}$} & \multicolumn{1}{l|}{$\Phi ^{+}$} & $\Phi
^{-}$ \\ \hline
\multicolumn{1}{|l|}{$Ancilla$ $3$} & \multicolumn{1}{l|}{$|1\rangle $} & 
\multicolumn{1}{l|}{$|1\rangle $} & \multicolumn{1}{l|}{$|0\rangle $} & $%
|0\rangle $ \\ \hline
\multicolumn{1}{|l|}{$Ancilla$ $4$} & \multicolumn{1}{l|}{$|0\rangle $} & 
\multicolumn{1}{l|}{$|1\rangle $} & \multicolumn{1}{l|}{$|0\rangle $} & $%
|1\rangle $ \\ \hline
\end{tabular}
\]

(4) Just as in step 2, after another pair of Hadamard gates , the final
output state recovers to the initial Bell state. So this measurement is a
quantum non-demolition process.

When a state $\Phi _{12}=a\Phi ^{+}+b\Phi ^{-}+c\Psi ^{+}+d\Psi ^{-},$where $%
a,b,c,d\in C$ , is inputted into the logic network shown in Fig.1, the
measurement process is just the same 
\begin{eqnarray}
&&(a\Phi ^{+}+b\Phi ^{-}+c\Psi ^{+}+d\Psi ^{-})\otimes |0\rangle _3\otimes
|0\rangle _4  \label{6} \\
&&\stackrel{(1)}{\rightarrow }(a\Phi ^{+}+b\Phi ^{-})\otimes |0\rangle
_3\otimes |0\rangle _4+(c\Psi ^{+}+d\Psi ^{-})\otimes |1\rangle _3\otimes
|0\rangle _4  \nonumber \\
&&\stackrel{(2)}{\rightarrow }(a\Phi ^{+}+b\Psi ^{+})\otimes |0\rangle
_3\otimes |0\rangle _4+(c\Phi ^{-}-d\Psi ^{-})\otimes |1\rangle _3\otimes
|0\rangle _4  \nonumber \\
&&\stackrel{(3)}{\rightarrow }a\Phi ^{+}|0\rangle _3\otimes |0\rangle
_4+b\Psi ^{+}|0\rangle _3\otimes |1\rangle _4+c\Phi ^{-}|1\rangle _3\otimes
|0\rangle _4-d\Psi ^{-}|1\rangle _3\otimes |1\rangle _4  \nonumber \\
&&\stackrel{(4)}{\rightarrow }a\Phi ^{+}|0\rangle _3\otimes |0\rangle
_4+b\Phi ^{-}|0\rangle _3\otimes |1\rangle _4+c\Psi ^{+}|1\rangle _3\otimes
|0\rangle _4+d\Psi ^{-}|1\rangle _3\otimes |1\rangle _4.  \nonumber
\end{eqnarray}
From the states of the two ancillary qubits, we could know which Bell state
has been measured out just as shown in Table 1.

Now, QND measurement of the Bell bases is completed. When we analyze the
measurement process, we see that controlled-not gates are employed to
extract the parity bit information. What is more critical is that Hadamard
gates can translate the phase qubit information QND Measurement to parity.
Since the Bell states include only one parity qubit and one phase qubit
information, two ancillary qubits are enough. And it can be seen easily that
in this measurement process, the information of the global phase of Bell
state are erased for only $\left| a\right| ^2,$ $\left| b\right| ^2,$ $%
\left| c\right| ^2,$ $\left| d\right| ^2$ can be known as the measurement
probability of each Bell states. This means that this phase operator
incommutes with the Bell operator.

\section{The generalization to N-Partite GHZ Bases}

For a $n$-partite generalized GHZ base in the Hilbert space $H_2^{\otimes n}$%
, there are $n-1$ part-parity qubits information which are between the
border-upon partite, and one phase qubit information. Then $n$ ancillary
qubits are needed. The key to this scheme is that controlled-not gates could
extract parity qubit information and this qubit phase information could be
translated into the holistic-parity information under Hadamard gates: 
\begin{eqnarray}
&&|\Phi ^{\pm }(x_0,x_1,...,x_{n-1})\rangle   \label{7} \\
&=&\Pi _{i=0}^{n-1}\otimes \hat{H}_i|\Psi ^{\pm
}(x_0,x_1,...,x_{n-1})\rangle   \nonumber \\
&=&\Pi _{i=0}^{n-1}\frac 1{\sqrt{2}}(|1\rangle +(-1)^{x_i}|0\rangle )\pm \Pi
_{i=0}^{n-1}\frac 1{\sqrt{2}}(|1\rangle -(-1)^{x_i}|0\rangle )  \nonumber \\
&=&\frac 1{2^n}\Sigma _{m=0}^{n-1}\Sigma
_{\{n_1,n_{2,}...,n_m\}}(-1)^{\Sigma _{l=1}^mx_{n_l}}(1\pm
(-1)^m)|\{n_{1,}n_2,...n_m\}\rangle ,  \nonumber
\end{eqnarray}
where $\{n_{1,}n_2,...n_m\}$ is a subset of $\{0,1,...,n-1\}$ and $\left|
\{n_{1,}n_2,...n_m\}\right\rangle $ denotes a $n$-partite state of which
only the $n_lth$ $(l=1,2,...,m)$ partite are $|1\rangle $.So it is easy to
see that each term in $|\Phi ^{+}(x_0,x_1,...,x_{n-1})\rangle $ contains an
even number of $|1\rangle $,while each term in $|\Phi
^{-}(x_0,x_1,...,x_{n-1})\rangle $ contains an odd number of $|1\rangle $.
That means the phase qubit information of $|\Psi ^{\pm
}(x_0,x_1,...,x_{n-1})\rangle $ have been translated into the $|\Phi ^{\pm
}(x_0,x_1,...,x_{n-1})\rangle $ global-parity information. So with $n-1$
part-parity qubits information and one global-parity qubit information
translated from the phase qubit, we could extract all the information
encoded in any $n$-partite GHZ bases. Obviously $n$ ancillary qubits which
are set in $|0\rangle $ at the beginning are needed. The process for quantum
non-demolition (QND) measurement of $n$-partite GHZ bases is similar, and
the logic network for GHZ states is shown in Fig. 2.

\section{Applications and Conclusion}

Since entanglement is introduced and preserved in the QND measurement
process, many novel applications, such as quantum authentication that
involves both features of the nonlocality and QND measurement, can be
exploited. Fig. 3 depicts a quantum authentication system in the Quantum
Security Automatic Tell Machine (QSATM) which employs just a logic network
of Bell states QND measurement. This authentication system can be used
circularly. From the figure, we can see that one of the input of the network
--- particle $2$, and the ancilla, particle $3$ and $4$ are all kept in the
QSATM. The other input, particle $1$, are stored in the QSATM credit card.
Particle $1$ and $2$ are set in one of the four Bell states, particle $3$
and $4$ are reset to $\left| 0\right\rangle $ every time before measurement.
There are two bits $3^{\prime }$ and $4^{\prime }$ in the disposer which
record the Bell state particle $1$ and $2$ originally in.

When a user inserts the credit card into the QSATM, the logic network works,
then particle $3$ and $4$ are set to states which indicates which Bell state
the output of particle $1$ and $2$ are in. \/If these states of particle $3$
and $4$ are the same as those recorded in bit $3^{\prime }$ and $4^{\prime }$%
, then this credit card is verified as legal, otherwise illegal. It is
evident that there is only $\frac 14$ probability for illegal user to pass
this verification process. Thus if $n$ pairs of Bell states are employed in
this system, then the probability of any illegal user passing the checkout
process is $(\frac 14)^n$. When $n$ is large enough, this probability will
approach $0.$

In practice, all the users whose authentication accuracy is higher than some
definite value are considered to be legal. In this case, the new states of
particle $3$ and $4$ are endowed to particle $3^{\prime }$ and $4^{\prime }$
which indicate the Bell state particle $1$ and $2$ are in now. It means that
as long as the decoherence of these Bell states is not too serious, the
system can be reset as particle $1$ and $2$ are set in a new Bell state
again. As illegal user can destroy the state of particle $2$ and thus all
the verification system. It is required that only those users who have the
right classical passwords have the access to quantum verification system.
But anyway, only those users who pass the quantum verification system are
believed to be legal, so it is much more reliable than the classical system.
It is guaranteed by the basic features of the quantum mechanics.

In conclusion, we have generalized the QND variables to the nonlocal ones by
proposing quantum non-demolition measurement networks for Bell states and
multi-particle GHZ bases with the help of controlled-not gates and Hadamard
gates which may be implemented on account of the recent progress in the
non-linear quantum optical\cite{r14,r15,r16,r17}. One of its applications in
the quantum authentication is also proposed.

\section{Acknowledgment}

This work was supported by the National Natural Science Foundation of China.

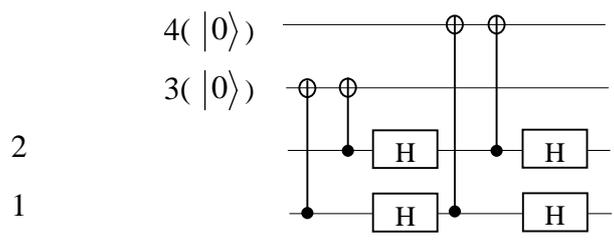
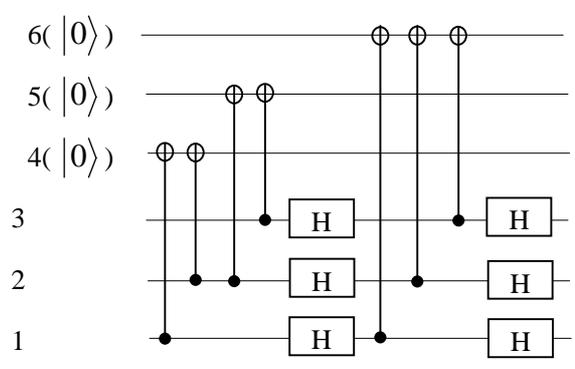
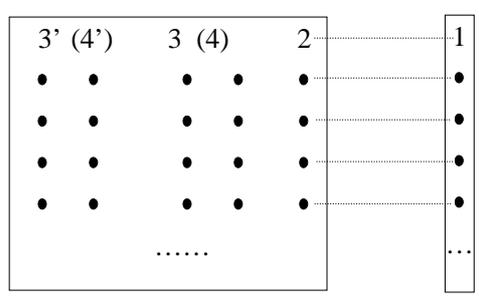
\end{document}